\documentclass[10pt,prd,twocolumn,aps,nofootinbib,nobibnotes,superscriptaddress,preprintnumbers]{revtex4-2}
\usepackage{graphicx,bm,color,amsmath,amssymb, mathtools,subcaption,multirow}
\graphicspath{{./fig/}}
\usepackage[hidelinks]{hyperref}
\hypersetup{
  colorlinks   = true, %Colours links instead of ugly boxes
  urlcolor     = [RGB]{46,48,146}, %Colour for external hyperlinks
  linkcolor    = [RGB]{46,48,146}, %Colour of internal links
  citecolor    = [RGB]{46,48,146} %Colour of citations
}

\usepackage{braket,orcidlink}

\usepackage{relsize}

\usepackage[font=small]{caption}

\newcommand{\EQ}{\begin{equation}}
\newcommand{\EN}{\end{equation}}
\newcommand{\EQA}{\begin{eqnarray}}
\newcommand{\ENA}{\end{eqnarray}}

{}
{}
{}

\def\OMG  W{{\Omega}_{\rm GW}}

\newcommand{\eV}{\,{\rm eV}}

\newcommand{\Hz}{\,{\rm Hz}}

\begin{document}

\title{Do Pulsar Timing Datasets Favor Massive Gravity?}

\date{\today}
\author{Chris~Choi\,\orcidlink{0009-0005-2328-3044}}
\email{Contact author: minyeonc@andrew.cmu.edu}
\affiliation{McWilliams Center for Cosmology and Astrophysics and Department of Physics, \href{https://ror.org/05x2bcf33}{Carnegie Mellon University}, Pittsburgh, Pennsylvania 15213, USA}

\author{Tina~Kahniashvili\,\orcidlink{0000-0003-0217-9852}}
\email{Contact author: tinatin@andrew.cmu.edu}
\affiliation{McWilliams Center for Cosmology and Astrophysics and Department of Physics, \href{https://ror.org/05x2bcf33}{Carnegie Mellon University}, Pittsburgh, Pennsylvania 15213, USA}
\affiliation{School of Natural Sciences and Medicine, \href{https://ror.org/051qn8h41}{Ilia State University}, 0194 Tbilisi, Georgia}
\affiliation{Department of Theoretical Astrophysics and Cosmology, \href{https://ror.org/02gkgrd84}{Georgian National Astrophysical Observatory}, Tbilisi, 47/57 M. Kostava St., GE-0179, Georgia}

\begin{abstract}

Several observational phenomena suggest that the standard model of cosmology and particle physics requires revision. To address this, we consider the extension of general relativity known as massive gravity (MG). In this Letter, we explore the imprints of MG on the propagation of gravitational waves (GWs): their modified dispersion relation and their additional (two vector and one scalar) polarization modes on the stochastic GW background (SGWB) detected by pulsar timing arrays (PTAs). We analyze the effects of massive GWs on the Hellings-Downs curve induced by modification of the overlap reduction function. Our study consists of analyzing observational data from the NANOGrav 15-year dataset and the Chinese PTA Data Release I, and is independent of the origin of the SGWB (astrophysical or cosmological). By considering the bound on the graviton mass imposed through the dispersion relation, we scrutinize the possibility of detecting traces of MG in the PTA observational data. We find that massive GWs predict better fits for the observed pulsar correlations. Future PTA missions with more precise data will hopefully be able to detect the GW additional polarization modes and might be effectively used to constrain the graviton mass.
\end{abstract}

\maketitle

\textit{Introduction}---The standard cosmological concordance model assumes that general relativity (GR) is the correct theory of gravity on cosmological length and time scales, and that the acceleration of the universe's expansion is due to a cosmological constant ($\Lambda$) that has a time-independent energy density and becomes dominant at late times \cite{Dodelson:2020bqr}. 
Despite the success of the concordance model, there are several unanswered questions, including the gauge hierarchy problem, the smallness of gravity compared to the other standard model forces, and the failure to formulate a unified theory of quantum gravity, among many others \cite{Dvali:2013qwe, Moffat:1998vi}. 
These puzzles suggest that GR is not suitable for describing physical processes at cosmological scales, and thus, one possibility is to assume that the true theory of gravity differs from GR \cite{deRham:2023byw}. 

One of the most active areas of research in gravity theory is in massive gravity (MG), which stems from the assumption that the graviton has a non-zero mass $m_g$. It may seem at first like a strange assumption, but there is no reason why the mass of the force carrier of gravity must be zero. The idea of a nonzero graviton mass has a long history, starting from the formulation of MG at the linear level by Fierz and Pauli \cite{Fierz:1939ix} in the 1930s. In contrast to GR, where gravitational waves (GWs) possess just two degrees of freedom with helicity (spin) $\pm 2$, tensor modes, generic MG theories have an additional three degrees of freedom, namely the helicity  $\pm 1$ vector and $0$ scalar modes. As a result, MG is subject to van Dam-Veltman-Zakharov (vDVZ) discontinuity \cite{vanDam:1970vg,Zakharov:1970cc}, where the five modes fail to reduce to the two tensor modes that we expect in GR in the massless limit. Taking into account the nonlinear effects of the strong gravitational potential, the Vainshtein mechanism \cite{Vainshtein:1972sx} is able to do away with the additional vector and scalar modes, freeing MG from the vDVZ discontinuity\footnote{GR is recovered in a strong gravitational field \cite{Tasinato:2013rza}, allowing us to verify GR in terrestrial and solar system level tests.}.

The effects of MG appear only on cosmological scales, possibly leading to an accelerated expansion \cite{DAmico:2011eto}. However, extensions in the nonlinear regime are unsatisfactory due to the presence of an unhealthy sixth ``ghost'' mode\footnote{Until 2010, it was thought that all Lorentz-invariant MG theories were characterized by the unhealthy presence of the ghost-mode, and thus were not valid \cite{deRham:2010kj}.} \cite{Boulware:1972yco}. More recently, groundbreaking progress has been made through the formulation of ghost-free MG in the de Rham-Gabadadze-Tolley (dRGT) theory \cite{deRham:2010ik,deRham:2010kj}, and its generalization to bigravity \cite{Hassan:2011zd} (see Ref.~\cite{deRham:2023ngf} for a review). Since then, many different modifications of dRGT and bigravity have been proposed \cite{Hinterbichler:2011tt,deRham:2014zqa,Koyama:2015vza,deRham:2016nuf,Hinterbichler:2016try, Cusin:2016ytz, Kenna-Allison:2019tbu, Kazempour:2022giy,Hogas:2021saw,Hogas:2021lns,Hogas:2021fmr,Hogas:2019ywm,Gialamas:2025ciw,Gialamas:2023fly,Gialamas:2023lxj,Gialamas:2023aim,Dwivedi:2024okk},  
resulting in the analysis of the consequences of MG and massive cosmologies (pioneered by Ref.~\cite{DAmico:2011eto} and further explored by Refs.~\cite{Gratia:2012wt,Gumrukcuoglu:2012aa,Maeda:2013bha,Akrami:2013pna,Zhang:2013noa,Lambiase:2012fv,Koyama:2011wx,Tasinato:2012ze,Solomon:2014iwa,Akrami:2013ffa,Koennig:2014ods,Gumrukcuoglu:2016hic, Heisenberg:2024uwq, Smirnov:2025yru, Comelli:2013tja}),
such as in the context of the creation of GWs and their propagation \cite{DeFelice:2013awa,Gumrukcuoglu:2013nza,DeFelice:2013bxa,DeFelice:2015moy,Babichev:2015xha,Sakstein:2017bws}. 
The direct detection of GWs at high frequencies by interferometers \cite{LIGOScientific:2016aoc,LIGOScientific:2016sjg, KAGRA:2020agh, VIRGO:2014yos} and the more recent detection of the stochastic GW background (SGWB) through various PTA collaborations \cite{Agazie:2023, Xu:2023wog,EPTA:2023sfo,EPTA:2023akd,EPTA:2023fyk, Zic:2023gta,Reardon:2023gzh} have ignited further interest in GW astronomy, giving us the unique possibility of constraining fundamental physics and understanding the nature of gravity. 
In our study, we used data collected by two GW detection missions, namely NANOGrav \cite{Agazie:2023} and Chinese PTA (CPTA) \cite{Xu:2023wog}, chosen for their unique approach to their data, to help us better understand gravity and connect MG to observables. 

In this Letter, we derive the overlap reduction function (ORF) and the dispersion relation in the case of ghost-free MG, taking into account the effect of these extra polarization modes. 
Unlike Refs.~\cite{Liang:2021bct, Anholm:2008wy, Arjona:2024cex, Lee:2013awh, Bernardo:2023mxc, Bernardo:2023pwt, Bernardo:2023zna, Bernardo:2024bdc}, we place an emphasis on the non-suppression of the exponential factors in the ORF, made possible by analyzing them within the context of an optimistic PTA observation prospect. In what follows, we use natural units, setting $c = $$\ \hbar = $$\ k_B = $$\ 1$.

\textit{Massive gravity}---The following effects of MG on GWs can be used to place constraints on the graviton mass \cite{deRham:2016nuf}:

(i) \textit{Yukawa-like exponential suppression}---Yukawa suppression is expected when the force carrier possesses a nonzero mass \cite{Will:1997bb}. In the case of MG, suppression of the gravitational potential is of the form $\Phi \propto e^{-m_gR}/R$ \cite{Will:1997bb}, dependent on the length scale $R$. There is also a suppression of GWs at wavelengths larger than the Compton length scale of the graviton ($R_g \simeq m_g^{-1}$)\footnote{The length scale determined by the graviton mass through the Yukawa suppression can naively be treated as a ``gravitational causal horizon'', outside of which we expect the correlations of gravitational perturbations to be suppressed \cite{Will:1997bb}.}.

(ii) \textit{modification of the dispersion relation}---The dispersion relation gains a massive term, implying a difference between the GW propagation speed ($c_g$) and the speed of light, which can be used to constrain $m_g$ \cite{LIGOScientific:2017vwq, LIGOScientific:2017zic, LIGOScientific:2017ync}. 

(iii) \textit{additional degrees of polarization}---These are characterized by the additional helicity $\pm 1$ and $0$ modes. Certain theories of MG, such as the minimal theory of MG \cite{DeFelice:2015hla}, only have two tensor modes, as in GR, but a general theory of MG that assumes Lorentz invariance will have five in total \cite{Comelli:2013tja}.

A resounding confirmation of the graviton's non-zero mass will involve the observation of all three effects at the corresponding scales. In this Letter, we turn our attention to the latter two effects, (ii) and (iii). Whenever the region of observation is much smaller than the radius of curvature induced by the GW, we may express the metric perturbation $h_{\mu\nu}$ as a plane wave \cite{Isi:2018miq}
\begin{equation}\label{eqn:planewave}
    h_{\mu\nu}(x) = \frac{1}{(2\pi)^4}\int d\boldsymbol{k} \ h_{\mu\nu}(k) e^{ikx} \ .
\end{equation}
Here, we have the 4-dimensional measure $d\boldsymbol{k} \equiv d^4 k 2\delta(|\boldsymbol{k}|^2 - |k_{\omega}|^2)/|\boldsymbol{k}|$ \cite{Isi:2018miq}, where $k_{\omega} \equiv \boldsymbol{k}(\omega)$ and is given by the dispersion relation. The dispersion relation can be derived directly from the relativistic energy-momentum relation, $E^2 = m_g^2 + |{\boldsymbol{p}}|^2$, by noting that $E = \omega$, the angular frequency of the GW, and ${\boldsymbol{p}} = \boldsymbol{k}$, the angular spatial wavevector of the GW. This yields the following expression for the dispersion relation: 
\begin{equation}\label{eq:dispersion}
    \omega^2 = |\boldsymbol{k}|^2+ m_g^2 \ ,
\end{equation}
where $\omega \equiv k_0 = 2\pi f$, the frequency of the GW. If the graviton were massless, then we would have the standard dispersion relation, $\omega^2 = |\boldsymbol{k}|^2$, but the dispersion relation gains nontriviality due to the non-zero mass in MG. 
To build intuition for this, we remind ourselves that an identical dispersion relation is obtained from the Proca equations \cite{Proca:1936fbw}, which describe massive helicity $\pm 1$ particles. A careful derivation shows how the mass term emerges \cite{Wang:2024kir} for the massive photon. It is therefore not unexpected that the dispersion relation for the massive graviton, in the context of a helicity $\pm 2$ field, takes on the same form. 

In general, the dispersion relation is dependent on the scale factor $a$ and the propagation speed of GWs $c_g$ \cite{Gumrukcuoglu:2012wt}. We assume that $a=1$, since we observe the GWs at present, and $c_g \sim  1$, based on observational constraints \cite{LIGOScientific:2017vwq, LIGOScientific:2017zic, LIGOScientific:2017ync}.
The dispersion relation provides a lower bound on the frequency for a GW, given by the mass of the graviton $m_g$, provided that $k$ goes to 0. In other words, $\omega$ must be greater than or equal to $m_g$; a departure from the behavior of GWs in GR.

\textit{Overlap reduction function}---Suppose that there is a massive GW propagating in an arbitrary spatial direction specified by the unit vector $\hat{\Omega} = (\sin\theta \cos\varphi,
                        \sin\theta \sin\varphi,
                        \cos\theta)$.
This GW affects PTA measurements by perturbing the spacetime along the line of sight of the pulsar signals, originally emitted with a constant frequency $\nu_0$, as they travel from their source pulsars to our telescopes. In the barycenter reference frame of the solar system, the frequency shift is characterized by the redshift $z(t) = (\nu_0 - \nu(t))/\nu_0$. PTA collaborations directly measure the time of arrival of pulsar signals, which corresponds to the integral of $z(t)$ \cite{Anholm:2008wy, Dahal:2020}, otherwise known as the anomalous residual, but it is trivial to rewrite these equations in terms of the residual if we so desire, so we proceed with the redshift. For PTA data, a quantity of interest is the two-point correlation of the total redshift, $\langle \tilde{z}(f) \tilde{z}(f') \rangle$, where $\tilde{z}(f)$ is the sky-averaged redshift $\tilde{z}(f) \equiv \int d^2\hat{\bf \Omega} z(f, \hat{\bf \Omega})$. Since the GWs are part of a stochastic ensemble, the SGWB, the relevant observable will be an angular correlation of the signals. The expression for the two-point correlation is given by \cite{Liang:2021bct, Anholm:2008wy}
\begin{equation}\label{eqn:two_point_z}
    \langle \tilde{z}(f) \tilde{z}(f') \rangle = \frac{3H_0^2\delta^2\left(\hat{\bf \Omega} , \hat{\bf \Omega}'\right)\delta_{ii'}\delta(f - f')}{32 \beta \pi^3|f|^3}\Omega_{\text{gw}}(|f|) \Gamma(|f|).
\end{equation}
Here, $\beta$ is the normalization factor introduced so that $\Gamma(|f|) = \frac{1}{2}$ for coaligned coincident pulsars to match the conventions of the Hellings-Downs (HD) correlation \cite{Romano:2023zhb}. $\Omega_{\text{gw}}(|f|)$ is the energy density of the GWs. Details of this derivation can be found in Refs.~\cite{Anholm:2008wy, Liang:2021bct}, which ultimately derive from even earlier work \cite{Detweiler:1979wn, Estabrook:1975jtn, Kaufmann:1970}. A wealth of literature has been produced to analyze the behavior of $\Omega_{\text{gw}}(|f|)$ in MG in the context of PTAs \cite{Choi:2023tun, Wu:2023rib, Kenjale:2024rsc, He:2021bqm}, but we turn our attention to the last factor in the expression: $\Gamma(|f|)$, otherwise known as the ORF. 

We assume that there are two pulsars, located at distances $L_1$ and $L_2$ from the solar-system barycenter, and in directions $\hat{p}_1$ and $\hat{p}_2$. For convenience, we assume pulsar 1 is situated along the $z$-axis and pulsar 2 is situated on the $x-z$ plane at an angle $\xi$ away from pulsar 1. We define the ORF for each polarization type: 
\begin{equation}\label{eq:orf_type}
    \Gamma_I(|f|) = \mathlarger{\sum}_{i \in I} \int_{S^2} d^2 \hat{\bf{\Omega}} \mathcal{E}_1(-f, \hat{\Omega}) \mathcal{E}_2(f, \hat{\Omega}) F_1^{(i)}(\hat{\bf{\Omega}}) F_2^{(i)}(\hat{\bf{\Omega}}) \ ,
\end{equation}
where $I = \{T,V,S\}$ for each type of polarization and we have defined the receiving function $F^{(i)}_j(\hat{\bf \Omega})$ and the exponential factor $\mathcal{E}_j(f, \hat{\Omega})$ for the $j$-th pulsar as 
\begin{equation}\label{eqn:recieving}
    \begin{aligned}
        F^{(i)}_j(\hat{\Omega}) &= \frac{\hat{p}^\mu_j}{2}\left(-\frac{\hat{p}^\nu_j}{1+\frac{|\boldsymbol{k}|}{k_0} \hat{\boldsymbol{\Omega}} \cdot \hat{{\boldsymbol{p}}_j}} \epsilon_{\mu \nu}^{(i)}+\epsilon_{0 \mu}^{(i)}\right) , \\ 
        \mathcal{E}_j(f, \hat{\Omega}) &= e^{-i2\pi f L_j\left( 1 + \frac{|\boldsymbol{k}|}{k_0} \hat{\bf \Omega}\cdot \hat{\boldsymbol{p}}_j\right)} - 1 \ . 
    \end{aligned}
\end{equation}
where $\epsilon_{\mu\nu}^{(i)}$ are the polarization tensors of helicity $i$ \cite{Liang:2021bct}. The scalar product $\hat{\bf \Omega}\cdot \hat{\boldsymbol{p}}_j$ comes directly from applying the delta function to the complex exponential in 
Eq.~(\ref{eqn:planewave}). 
It is possible to decompose the ORF into different modes because the tensor, vector, and scalar modes decouple from each other with an appropriate gauge-fixing term, allowing the kinetic terms to fully diagonalize \cite{Hinterbichler:2011tt}. Our effective overlap reduction function is given by 
\begin{equation}\label{eq:eff_orf}
    \tilde{\Gamma}_{T} = \beta \left(\Gamma_{T} + \Gamma_{V} \frac{\Omega_V}{\Omega_T} + \Gamma_{S} \frac{\Omega_S}{\Omega_T} \right) \ ,
\end{equation}
which behaves like an effective tensor mode, since PTA observations are not capable of distinguishing the different polarization modes from each other \cite{Liang:2021bct}. In general, the energy densities $\Omega_T, \Omega_V,\Omega_S$ are frequency dependent, but we suppress the frequency dependence in our analysis and assume that the energy density is equipartitioned, that is, $\Omega_T = \Omega_V = 2\Omega_S$.

\textit{Observational constraints}---Pulsar timing arrays are sensitive to frequencies in the range \cite{Moore:2014lga}
\begin{equation}\label{eq:freqrange}
    \frac{1}{T_{\text{obs}}} < f < \frac{1}{\delta t}\ ,
\end{equation} 
where $T_{\text{obs}}$ is the total length of time the pulsars are observed and $\delta t$ is the cadence, that is, how often the pulsars are observed. In the best-case scenario of PTA observation, we would be able to observe pulsars for nearly a century, if not longer. This corresponds to a frequency of $f_{\text{min}} \sim 3.17 \times 10^{-10} \Hz$, setting the lower limit of our frequency range. The closest pulsars used for PTAs are about $L_{\text{min}} \sim 100$ ly away \cite{Anholm:2008wy}, yielding $fL \sim 1$. Therefore, we do not suppress the exponential factors, unlike in Refs.~\cite{Liang:2021bct,Arjona:2024cex}, because $fL$ is well within the regime where the exponential factors are not negligible.
We used Monte-Carlo integration to numerically compute the frequency dependence of the effective ORF. The magnitude of $\Gamma_T(|f_{\text{min}}|)$ is significantly greater than $\Gamma_T$ with the factors of $\mathcal{E}(f, \hat{\bf \Omega})$ suppressed, so the approximation of $\mathcal{E}_1(-f)\mathcal{E}_2(f) \sim 1$ is not appropriate. If the ORF is in the regime where the Taylor expansions in the short wavelength ($fL \gg 1$) or long wavelength ($fL \ll 1$) cannot be carried out, then we must not ignore $\mathcal{E}(f, \hat{\bf \Omega})$.
\begin{figure}[ht]
    \centering
    \includegraphics[scale=0.5]{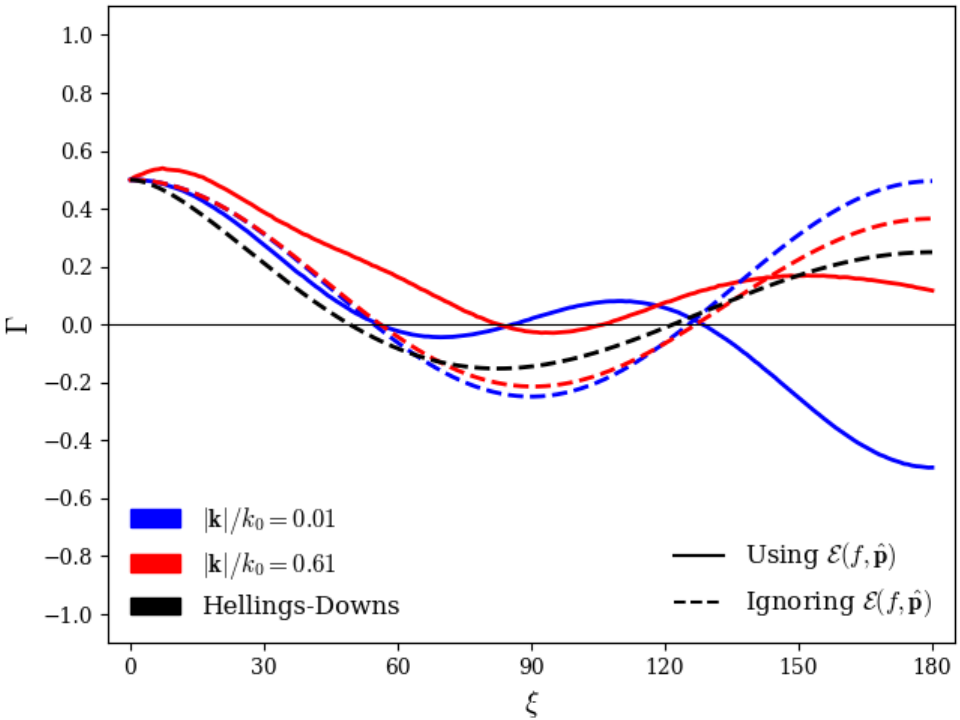}
    \caption{The ORFs plotted as a function of angular separation $\xi$. The solid lines use the full expression of the ORFs, including $\mathcal{E}$. The dotted lines assume $\mathcal{E}_1(-f)\mathcal{E}_2(f)=1$.}
    \label{fig:orfs}
\end{figure}
In Fig.~\ref{fig:orfs}, we see how differently the ORFs behave when we take $\mathcal{E}$ into account. We observe that for fixed $fL$, the ratio $|\boldsymbol{k}|/k_0$ corresponds to an increasing disparity between the ORFs with and without $\mathcal{E}$. For sufficiently small values of the ratio $|\boldsymbol{k}|/k_0$, e.g.\ $|\boldsymbol{k}|/k_0 \lesssim 0.1$, the ORFs with $\mathcal{E}$ acquire additional local extrema, namely a minimum and a maximum. This is a peculiarity that only arises when $\mathcal{E}$ is not ignored. We also expect this behavior to emerge when the mass of the graviton is sufficiently high.

The graviton mass is intimately connected to the behavior of the effective ORF. It appears through a relationship between the mass and the angular frequency. 
If observations of PTAs follow the trajectory for the best-case scenario, then we can only hope to observe a graviton mass given by that lower frequency bound. This corresponds to $m_g \sim 1.31\times 10^{-24} \eV$. To be clear, we are assuming that the lowest theoretical frequency given by the limit of the dispersion relation (\ref{eq:dispersion}) as $k\rightarrow 0$, will coincide with the lowest frequency that PTAs can measure, hence the ``best-case'' scenario. Going forth, this is the mass that we will take $m_g$ to be in our analysis.

\begin{figure*}[ht]
    \includegraphics[width=.95\textwidth]{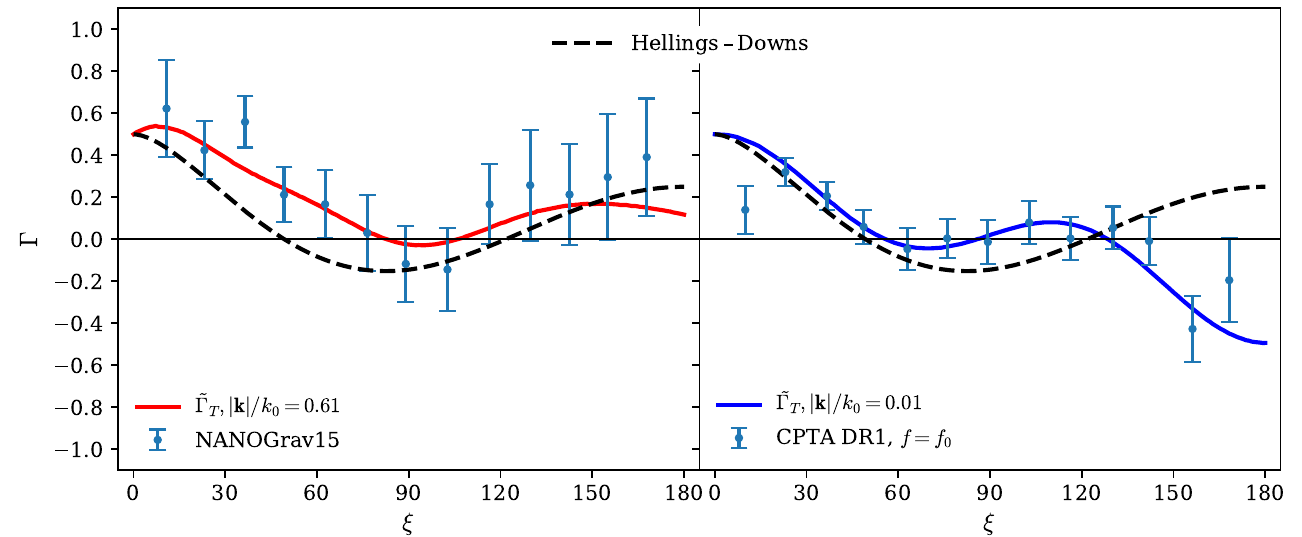}
    \caption{The effective ORF plotted as a function of the angular separation $\xi$ between a pair of pulsars, for $|\boldsymbol{k}|/k_0 = 0.61$ in red and $|\boldsymbol{k}|/k_0 = 0.01$ in blue. NANOGrav15 is plotted with error bars in the left panel, and CPTA DR1 in the right panel, for $f = 1/T_{\text{CPTA}}$. For both panels, we depict the HD correlation as a reference.}
    \label{fig:pta_compare}
\end{figure*}

The graviton mass affects all the ORFs for each polarization type, including the tensor modes. The graviton mass shows up in exponential terms as the coefficient in front of the dot product $\hat{\bf \Omega}\cdot \hat{\boldsymbol{p}}_j$ and in the massive spin-1 helicity-0 polarization vector $\epsilon_\mu^0(m_g)$. The former appears in all of the polarization types, but the latter only appears in the vector and scalar modes in $F_j^{(i)}$, further complicating their mass dependence. Interestingly, because the graviton mass appears only as a ratio $m_g / k_0$, the behavior of the ORF is only sensitive to the mass relative to the frequency. In other words, there is no significance for an ORF purely due to a graviton mass of $m_g$. The question would be, with regard to what frequency?

\textit{Results}---We now present the main results of this work. Although our calculations are for observations that extend well into the future, it is instructive to compare them to current data and confirm whether they are within the present standard deviations, or perhaps explain the current data better than HD. We compute $\tilde{\Gamma}_T$, keeping $m_g$ fixed and the ratio $|\boldsymbol{k}|/k_0$ as a free parameter. We then compare with the NANOGrav 15-year dataset (NANOGrav15) \cite{Agazie:2023} and with the CPTA Data Release I (DR1) \cite{Xu:2023wog}, as seen in Fig.~\ref{fig:pta_compare}.

We used NANOGrav15 to obtain angular correlations of 2,211 pairs of pulsars from a 67-pulsar array \cite{Agazie:2023}. We do not use the frequentist optimal statistic that Ref.~\cite{Agazie:2023} employs, which is based on methods described in Ref.~\cite{Allen:2022ksj}, since it is crucial that the estimator is not biased towards the HD curve. We do not want the statistics to assume anything about the additional polarizations. We used 13 bins constructed such that the mean of each bin matches the mean $\xi$'s of CPTA DR1 as closely as possible, since we are not able to alter the number or range of the bins in CPTA DR1 due to its public inaccessibility. As a result, we computed the statistics for each binned average rather than the correlation coefficient for each pair so that the two datasets are on equal footing. 

We find that for NANOGrav15, HD gives a fit of $\chi^2$/d.o.f.\ $\sim 1.71$, whereas the best-fit $\tilde{\Gamma}_T$ has a ratio $|\boldsymbol{k}|/k_0 = 0.61$, with $\chi^2$/d.o.f.\ $\sim 0.55$. The effective ORF from MG is therefore a better overall fit. A curious property of the effective ORF in MG is the shift of the minimum angle. The apparent minimum in NANOGrav15 is located further to the right than HD predicts: HD predicts $\xi \sim 82^\circ$ for HD and the data shows that $\xi \sim 97^\circ$, which we obtain using parabolic interpolation. The $\tilde{\Gamma}_T$ that we obtained has a minimum angle at $\xi \sim 95^\circ$.

We used CPTA DR1 to manually reconstruct the binned average values from 1,596 pairs of pulsars in a 57-pulsar array \cite{Xu:2023wog}. This data set is chosen due to the frequency dependence observed by CPTA \cite{Xu:2023wog}. Even in the absence of $\mathcal{E}$, we still expect a frequency dependence in the data for MG. CPTA analyzed the correlation coefficients for $f=$ $1/T_{\text{CPTA}}, 1.5/T_{\text{CPTA}},$ and $2/T_{\text{CPTA}}$, where $T_{\text{CPTA}} \sim 3.40$ yrs. We use the data for $1/T_{\text CPTA}$, since it diverges from HD to the greatest extent. 

\begin{figure}[ht]
    \centering
    \includegraphics[width=0.47\textwidth]{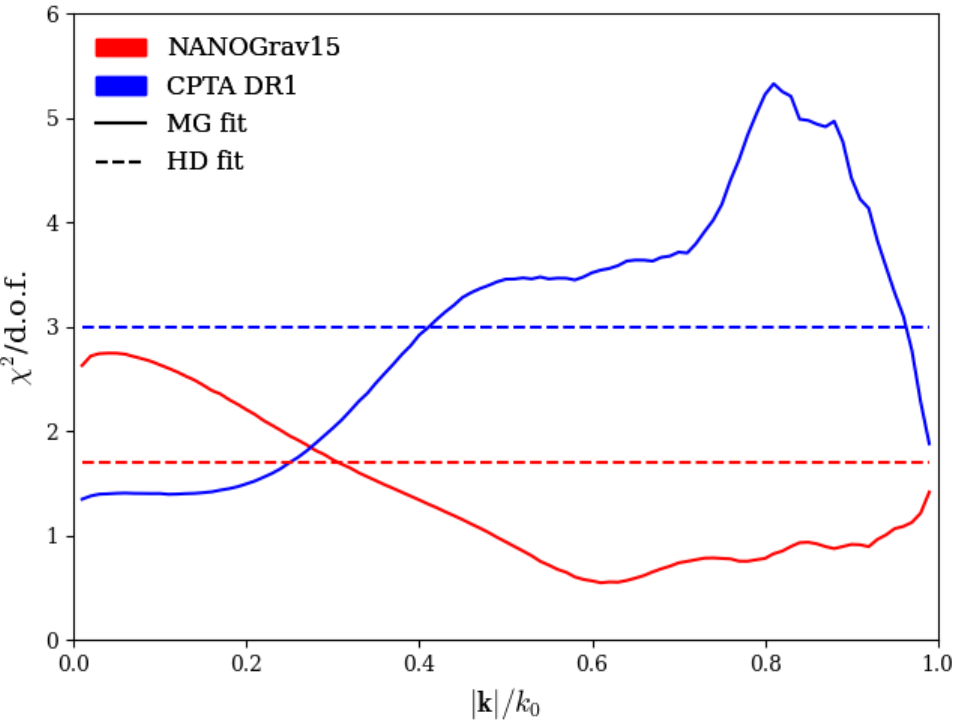}
    \caption{The $\chi^2$/d.o.f.~fitting plotted as a function of the ratio $|\boldsymbol{k}|/k_0$. The values for the horizontal dashed lines, representing the fit for the HD, can be found in Table \ref{tbl:chi}.}
    \label{fig:chi}
\end{figure}

We find that for CPTA DR1, HD gives a fit of $\chi^2$/d.o.f.\ $\sim 3.00$ whereas the best-fit $\tilde{\Gamma}_T$ has a ratio $|\boldsymbol{k}|/k_0 = 0.01$ with $\chi^2$/d.o.f.\ $\sim 1.35$, a markedly improved fit. The effective ORF from MG captures the behavior of CPTA DR1 remarkably better than HD; the HD correlation fails to capture both the apparent additional maximum angle, located at $\xi \sim 103^\circ$, and the shifted minimum angle, located at $\xi \sim 65^\circ$. Our best-fit $\tilde{\Gamma}_T$ has its maximum angle at $\xi \sim 108^\circ$ and minimum angle at $\xi \sim 70^\circ$. We summarize our statistics in Table \ref{tbl:chi} and the results for our $\chi^2$ fitting in Fig.~\ref{fig:chi}.
\begin{table}[ht]
\centering
\renewcommand{\arraystretch}{1.8}
\begin{tabular}{|c|c|c|c|c|}
\hline
\textbf{Data} & \textbf{Fit type} & \textbf{Best fit} $\frac{\boldsymbol{k}}{k_0}$ & \textbf{$\chi^2$} & \textbf{$\chi^2$/d.o.f.} \\
\hline
\multirow{2}{*}{NANOGrav15} & HD & \textbf{--}  & 22.20 & 1.71 \\ \cline{2-5}
                           & MG & 0.61 & 6.59  & 0.55 \\ 
\hline
\multirow{2}{*}{CPTA DR1}  & HD & \textbf{--} & 38.95 & 3.00 \\ \cline{2-5}
                           & MG & 0.01 & 16.58 & 1.35 \\ 
\hline
\end{tabular}
\caption{The $\chi^2$ and $\chi^2$/d.o.f.\ values for different fit functions for the two datasets used in this analysis. We have 13 degrees of freedom for the HD correlation and 12 for the MG model (one fit parameter, $|\boldsymbol{k}|/k_0$).}
\label{tbl:chi}
\end{table}

\textit{Discussion}---In this paper, we have reviewed the methods to obtain the modified dispersion relation and the effective ORF in the theory of ghost-free MG. We have analyzed the behavior of this effective ORF when $\mathcal{E}$ is not ignored. We were able to demonstrate that using unbiased NANOGrav15 and CPTA DR1, the effective ORF in MG fits the data significantly better than the HD curve with an optimistic graviton mass $m_g \sim$ $1.31\times10^{-24} \eV$.

A rigorous fitting to NANOGrav15 without considering $\mathcal{E}$ has been done in Ref.~\cite{Arjona:2024cex}, which found that the best-fit ratio to the frequentist-optimal statistic of NANOGrav15 is $|\boldsymbol{k}|/k_0 \sim 0.73$. The authors of Ref.~\cite{Arjona:2024cex} set $\Omega_V = \Omega_S$ and had $\Omega_T$ as the second fit parameter, obtaining a best‑fit value of 0.46 and a $\chi^2$ of 6.91, slightly worse than our 6.59. While we do observe tension between the PTA datasets and the HD interpretation, opening the door for beyond-GR theories such as MG to provide a better fit, it may be addressed by more data from higher precision measurements and extended observing campaigns, as well as improved modeling techniques. 

An assumption made in this Letter is the isotropy of the SGWB, allowing us to decompose Eq.~\ref{eqn:two_point_z} as we did, but it is not necessarily one we can take for granted \cite{Depta:2024ykq, Bravo:2025csu, Cusin:2025xle, Kuwahara:2024jiz, Li:2024lvt}. Anisotropy in the background can be analyzed by decomposing the ORF on the basis of spherical harmonics and analyzing multipole moments \cite{Allen:2024bnk, Gair:2014rwa}\footnote{It may not be appropriate to do spherical harmonic decomposition for the analysis of PTAs \cite{Ali-Haimoud:2020ozu}. Instead, the Fischer formalism may be more effective, especially in the context of anisotropic backgrounds.}.  
This can be applied to the effective ORFs we have derived and may be of interest in further work. 

Frequency-dependent analysis of the angular correlation of PTAs in general is underexplored. NANOGrav and other PTA collaborations do not claim to detect any frequency dependence in the ORF \cite{Agazie:2023,EPTA:2023sfo,EPTA:2023akd,EPTA:2023fyk, Zic:2023gta,Reardon:2023gzh}, while CPTA has \cite{Xu:2023wog}. If the graviton is massive, then there will certainly be a frequency dependence in the data, even without considering $\mathcal{E}$. Calculating the frequency dependence of the pulsar correlations in the rest of the PTA observations will be quite elucidative. 

These results do not depend on the origins of the SGWB. If gravitons are massive, then they will behave according to the principles presented in this Letter, and we should expect to observe the three effects of MG, regardless of whether the source is astrophysical or cosmological in origin. In both scenarios, the propagation of the GWs is not affected; properties such as the spectral shape and range of the amplitude \cite{Caprini:2015tfa} or the isotropy of the background \cite{NANOGrav:2023hvm} are affected.

The detection prospects suggested in this Letter are quite optimistic. It may be unlikely that a PTA experiment would last that long, but we nonetheless provide a detailed overview of the ORFs in such a case. With regard to data from current PTA collaborations, the prospects seem hopeful of continuing to detect pulsar-pulsar correlation coefficients such that the ORFs generated in MG stay within their standard deviations. With more pulsars being added to PTAs and the planned missions of space-based GW observatories on the horizon \cite{LISA:2017, TianQin:2015yph, Hu:2017mde}, it seems plausible that the mysteries that have evaded our investigations surrounding the cosmos may finally start to be unravelled. Clearly, a new era of GW observations is well underway. 

\vspace{3mm}
\textit{Acknowledgements}---We thank Qiuyue Liang, Neil Cornish, and Murman Gurgenidze for useful discussions related to the paper. We also thank the organizers of the 2025 Phenomenology Symposium (PHENO), during which much of this paper has been developed. C.C.~and T.K.~acknowledge support from the NASA Astrophysics Theory Program (ATP) Award 80NSSC22K0825 and the National Science Foundation (NSF) Astronomy and Astrophysics Research Grants (AAG) Award AST2408411. T.K.~acknowledges partial support from the Shota Rustaveli National Science Foundation of Georgia, FR24-901
\vspace{3mm}
\textit{Data availability}---Source code and data for reproducing all of our results are openly available \cite{Choi:2025git}.

\bibliographystyle{apsrev4-2_edited}
\bibliography{refs}

\clearpage
\end{document}